\documentclass[a4paper,UKenglish,cleveref, autoref, thm-restate,authorcolumns]{lipics-v2019}
\usepackage{algorithm}
\usepackage[algo2e,ruled,vlined,linesnumbered]{algorithm2e}
\usepackage{booktabs}
\newif\ifTR
\TRtrue

\usepackage{amsfonts}

\newcommand{\set}[1]{\left\{ #1\right\}}

\newcommand{\sodass}{\,:\,}
\newcommand{\setGilt}[2]{\left\{ #1\sodass #2\right\}}

\newcommand{\realrange}[2]{\left[#1, #2\right]}

\newcommand{\unitrange}[2]{\realrange{0}{1}}

\newcommand{\discussionsize}{\small}

\newcommand{\frage}[1]{}

\newsavebox{\codeparam}

{\end{disscodepos}}

\newcommand{\Is}       {:=}

\newdimen\endofsize\endofsize=0.5em
\def\endofbeweis{~\quad\hglue\hsize minus\hsize
                 \hbox{\vrule height \endofsize width
\endofsize}\par}

\usepackage{numprint}
\DeclareMathOperator\mate{\mathrm{mate}}

\newcommand{\ie}{i.e.~}
\newcommand{\etal}{et~al.~}
\newcommand{\eg}{e.g.}

\newcommand{\csch}[1]{{\color{blue}[CS: #1]}}
 \renewcommand{\csch}[1]{{}}

\ifTR
\nolinenumbers
\else \fi{}
\newcommand{\mytitle}{Dynamic Matching Algorithms in Practice}
\title{\mytitle}
\ifTR
\author{Monika Henzinger}{University of Vienna, Faculty of Computer Science, Vienna, Austria}{monika.henzinger@univie.ac.at}{https://orcid.org/0000-0002-5008-6530}{}
\author{Shahbaz Khan}{Department of Computer Science, University of Helsinki, Finland}{shahbaz.khan@helsinki.fi}{https://orcid.org/0000-0001-9352-0088}{}
\author{Richard Paul}{University of Vienna, Faculty of Computer Science, Vienna, Austria}{richard.paul@univie.ac.at}{https://orcid.org/0000-0002-7433-0075}{}
\author{Christian Schulz}{University of Vienna, Faculty of Computer Science, Vienna, Austria}{christian.schulz@univie.ac.at}{https://orcid.org/0000-0002-2823-3506}{}
\authorrunning{M. Henzinger, S. Khan, R. Paul, C. Schulz}
\fi{}
\hideLIPIcs
\ifTR
\funding{
  The research leading to these results has received funding from the European Research Council under the
European Community's Seventh Framework Programme (FP7/2007-2013) /ERC grant agreement No. 340506. }

\fi{}
\date{}

\begin{document}

\maketitle

\begin{abstract}

In recent years, significant advances have been made in the design and analysis of fully dynamic maximal matching algorithms.  However, these theoretical results have received very little attention from the practical perspective. Few of the algorithms are implemented and tested on real datasets, and their practical potential is far from understood.  In this paper, we attempt to bridge the gap between theory and practice that is currently observed for the \emph{fully dynamic maximal matching problem}. We engineer several algorithms and empirically study those algorithms on an extensive set of dynamic instances.
\end{abstract}
\thispagestyle{empty}
\section{Introduction}
{}
\newcommand{\EM}{${\cal E}_M$\xspace}
\newcommand{\Em}{{\cal E}_M}
The matching problem is one of the most prominently studied combinatorial graph problems having a variety of practical applications. 
A matching $\mathcal{M}$ of a graph $G=(V,E)$ is a subset of edges such that no two
elements of $\mathcal{M}$ have a common end point. Many applications require 
matchings with certain properties, like being maximal (no edge can
be added to $\mathcal{M}$ without violating the matching property) or having maximum
cardinality. These
problems can be solved in polynomial time.
For example, Micali and Vazirani~\cite{DBLP:conf/focs/MicaliV80} compute a maximum cardinality matching in $O(m\sqrt{n})$ time. For the weighted case, the fastest algorithm is by Galil et. al~\cite{GalilMG86} requiring $O(mn\log n)$ time which improves the $O(n^3)$ time algorithm~\cite{Gabow74} for sparse graphs. 

However, often the underlying graphs change over time, \eg, edges are inserted or deleted in the graph as the time progresses.
For example, new relations between objects of a network may be created or removed over time (for example \cite{DBLP:conf/focs/MehtaSVV05}).
Even though the matching problem can be solved in polynomial time, computing a new matching from scratch every time the graph changes is an expensive task on huge networks, as this ignores the previously computed information on the given instance.
Hence, in the recent years significant advances have been made in the design and analysis of fully dynamic maximal matching 
algorithms. These theoretical algorithmic ideas have received very little attention from the practical perspective. 
Only a few of the dynamic algorithms are implemented and tested on real datasets, and hence their practical potential is far from
being understood.

\textbf{Contribution and Outline.}
In this paper, we start to bridge the gap between theory and practice that is currently observed for the \emph{fully dynamic maximal matching problem}. 
We engineer several dynamic maximal matching algorithms as well as an algorithm that is able to maintain the maximum matching. 
To this end, we look at an algorithm due to Baswana, Gupta and Sen \cite{BaswanaGS15}, which performs edge updates in  $O(\sqrt{n})$ time and maintains a 2-approximate maximum matching, the algorithm of Neiman and Solomon~\cite{NeimanS16}, which takes $O(\sqrt{n+m})$ time to maintain a 3/2-approximate maximum matching, as well as two \emph{novel} dynamic algorithms: a random walk-based algorithm as well as a dynamic algorithm that searches for augmenting paths using a (depth bounded) blossom algorithm. Without depth bound, the latter algorithm is able to maintain a maximum matching.
We perform extensive experiments comparing the performance of these algorithms on the real-world and artificially generated instances. 
Experiments indicate that maintaining optimum matchings can be done much more efficiently than the naive algorithm that recomputes maximum matchings from scratch (more than an order of magnitude faster). Second, all non-optimum dynamic algorithms that we consider in this work are able to maintain near-optimum matchings in practice while being multiple orders of magnitudes faster than the naive optimum dynamic algorithm.

\section{Preliminaries}

\textbf{Basic Concepts.} Let  $G=(V=\{0,\ldots, n-1\},E)$ be an \emph{undirected graph} 
without parallel edges and self-loops. We set $n = |V|$, and $m = |E|$,
$N(v)\Is \setGilt{u}{\set{v,u}\in E}$ denotes the \emph{neighbors} of $v$.
The degree of a vertex $v$ is $d(v):=|N(v)|$.
A matching $\mathcal{M} \subset E$ in a graph is a set of edges without common vertices.
The \emph{cardinality} or \emph{size} of a matching is simply the cardinality of the edge subset $\mathcal{M}$.
We call a matching \emph{maximal}, if there is no edge in $E$ that can be added to $\mathcal{M}$. 
A \textit{maximum cardinality matching} $\mathcal{M}_{\text{opt}}$ is a matching that contains the largest possible
number of edges of all matchings.
An $\alpha$-\emph{approximate} maximum matching is a matching, that contains at least $\frac{|\mathcal{M}_{\text{opt}}|}{\alpha}$ edges.
A vertex is called \emph{free} or \emph{unmatched} if it is not incident to an edge of the matching. Otherwise, we call it \emph{unfree} or \emph{matched}.
For a matched vertex $u$ with $\{u,v\} \in \mathcal{M}$, we call vertex $v$ the \emph{mate} of $u$, which we denote as $\mate(u)=v$. 
For an unmatched vertex $u$, we define $\mate(u)=\bot$. 
An \emph{augmenting path} is defined as a cycle-free path in the graph $G$, that starts and ends on a \emph{free} vertex and where edges from $\mathcal{M}$ alternate with edges from $E \setminus \mathcal{M}$. The \emph{trivial augmenting path} is a single edge, that has both its endpoints unmatched. Throughout this paper, we call such an edge a \emph{free} edge. If we take an augmenting path and resolve it by matching every unmatched edge and unmatching every matched edge, we increase the cardinality of the matching  by one.
Any matching without \emph{augmenting paths} is a maximum matching~\cite{berge57} and  any matching with no augmenting paths of length at most $2k-3$ is a $(k/(k-1))$-approximate~maximum~matching~\cite{hopkarp71}. Hence, a maximal matching having no augmenting paths of length one (or free edges) is a $2$-approximate~maximum~matching. Throughout the paper, we omit the inverse Ackermann function from complexity statements.

Our focus in this paper are \emph{fully dynamic graphs}, where the number of vertices is fixed, but edges can be added and removed. 
All the algorithms evaluated can handle edge insertions as well as edge deletions.
In the following, $\Delta$ denotes the maximum degree that can be found in any state of the dynamic graph.

\noindent\textbf{Related Work.} 
\label{sec:relatedwork}
\ifTR
Computing large or maximum matchings in graphs is a well researched topic. Edmonds~\cite{edmonds1965paths} gave an algorithm that can compute a maximum cardinality matching in a static graph in time $O(mn^2)$. This result was later improved to $O(m n ^ {0.5})$ by Micali and Vazirani \cite{DBLP:conf/focs/MicaliV80}. Recently, algorithms use simple data reductions rules such as \cite{DBLP:conf/esa/KorenweinNNZ18} to speed up computations, or shrink-trees instead of blossoms~\cite{DBLP:conf/alenex/DroschinskyMT20} to speed up computations in static graphs. 
In practice, these algorithms can still be time consuming for many applications involving large graphs. 
Hence, several near
linear~time~approximation algorithms exist in practice such as the local max algorithm \cite{DBLP:conf/europar/BirnOSSS13}, the path growing algorithm~\cite{DH03a} and the global paths algorithm~\cite{MauSan07}. 
As the focus of this work are dynamic graphs, we refer the reader to the quite extensive related work section of~\cite{DBLP:conf/alenex/DroschinskyMT20} for more recent static matching algorithms.

\fi{}
In the dynamic setting, the maximum matching problem has been prominently studied ensuring $\alpha$-approximate guarantees. 
A major exception is the randomized algorithm by Sankowski \cite{Sankowski07} which maintains a maximum matching in $O(n^{1.495})$ update time. 
One can trivially maintain a maximal ($2$-approximate) matching in $O(n)$ update time by resolving all trivial augmenting paths of length one. 
Ivkovi\'c and Llyod \cite{IvkovicL93} designed the first fully dynamic algorithm to improve this bound to $O((n+m)^{\sqrt{2}/2})$ update time. 
Later, Onak and Rubinfeld~\cite{OnakRubinfeld10} presented a randomized algorithm for maintaining a $O(1)$-approximate matching in a dynamic graph that takes $O(\log^2 n)$ expected amortized time for each edge update. 
This result led to a flurry of results in this area. 
Baswana, Gupta and Sen~\cite{BaswanaGS15} improved the approximation ratio of \cite{OnakRubinfeld10} from $O(1)$ to $2$ and the amortized update time to $O(\log n)$. 
Further, Solomon~\cite{Solomon16} improved the update time of~\cite{BaswanaGS15} from amortized $O(\log n)$ to {\em constant}. 
However, the first deterministic data structure improving~\cite{IvkovicL93} was given by  Bhattacharya et al.~\cite{BhattacharyaHI18} maintaining $(3+\epsilon)$ approximate matching in $\tilde{O}(min(\sqrt{n},m^{1/3}/\epsilon^2))$ amortized update time, which was further improved to $(2+\epsilon)$ requiring $O(\log n/\epsilon^2)$ update time by Bhattacharya et al.~\cite{Bhattacharya2016}. 
Recently, Bhattacharya et al.~\cite{BhattacharyaCH17} achieved the first $O(1)$ amortized update time for a deterministic algorithm but for a weaker approximation guarantee of $O(1)$. 
For worst-case bounds, the best results are by Gupta and Peng~\cite{GuptaP13} requiring $O(\sqrt{m}/\epsilon)$ update time for $(1+\epsilon)$ approximation, Neiman and Solomon~\cite{NeimanS16} requiring $O(\sqrt{m})$ update time for $3/2$ approximation, Bernstein and Stein~\cite{Bernstein2016a} requiring $m^{1/4}/\epsilon^{2.5}$ for $(3/2+\epsilon)$ approximation. 
Recently, Charikar and Solomon~\cite{CharikarS18}, and Arar et al.~\cite{ArarCCSW18} (using \cite{Bhattacharya2017b}), independently presented the first algorithms requiring $O(poly\log n)$ worst-case update time both maintaining $(2+\epsilon)$ approximation.
Recently, Grandoni~\etal\cite{DBLP:conf/soda/0001LSSS19} gave an incremental matching algorithm that achieves a $(1+\epsilon)$-approximate matching in constant deterministic amortized time.

Despite this variety of different algorithms, to the best of our knowledge, there has been no effort made so far, to engineer and evaluate these algorithms on real-world instances. Moreover, although there exist quite numerous randomized algorithms for the dynamic maximal matching problem, we do not know about any attempts, to use random walks as a mean to improve matching quality.
\section{Algorithms}
\label{sec:algorithms}

We now present the fully dynamic algorithms for the maximal matching problem under consideration. 
We implemented and tested a variety of simple, combinatorial algorithms that
seemed likely to work well in practice.
We begin with random walk based dynamic algorithms, followed by dynamic algorithms based on (bounded) augmenting path search and finally review  the algorithms by Baswana, Gupta and Sen~\cite{BaswanaGS15} and Neiman and Solomon~\cite{NeimanS16}.
All of the algorithms are fully dynamic. 
 Throughout this section, we provide a brief description of these algorithms and their implementation. {}

\subsection{Random Walk-based Algorithms}
\label{sec:random-walk}
In general finding long augmenting paths in networks is an expensive step.
The main idea of the random walk based methods proposed in this section is to use random walks in order to detect augmenting paths, and hence to improve the size of the matching. 
We start by explaining the general idea to use random walks for finding augmenting paths and then explain how we handle edge insertions and deletions.

\subsubsection{Random Walks For Augmenting Paths}
The algorithm works as follows: 
we start at a free vertex $u$ and randomly choose a neighbour $w$ of $u$. 
If this neighbour is free, then we match the edge $(u,w)$ and our random walk stops. 
If $w$ is matched, we unmatch $(w, \mate(w))$ and match $(u,w)$. Note that $u \neq \mate(w)$ since $u$ is free in the beginning and therefore $\mate(u) = \bot$, but $\mate(\mate(w)) = w$ and $w \neq \bot$. Afterwards, the previous mate of $w$ is free. Hence, we continue our random walk at this vertex. 
Our random walk performs $O(\frac{1}{\epsilon})$ steps (see below). 
Since picking a random neighbor can be done in constant time, the overall time for the random walk update algorithm is $O(\frac{1}{\epsilon})$.
Note that the length of the random walk is a natural parameter of the algorithm that we will investigate in the~experimental~evaluation.

Also note that if the algorithm does not end by matching two free vertices, the matching may not be maximal even if it was initially -- this can be the case if the vertex freed last is incident to a free vertex. 
There are multiple possibilities to fix this. 
Our default is to undo all changes that~have~been~done in this case. The overall running time of a random walk is then $O(1/\epsilon)$.
Another possibility is \textit{$\Delta$-settling:} 
The algorithm tries to settle visited vertices. The algorithm scans through their neighbors to find a free vertex and stops if once it was successful or the number of steps exceeds $1/\epsilon$. If the random walk was not successful, the algorithm tries to match the last vertex touched by the random walk by scanning its neighbors instead of undoing all changes. 
This also ensures that the matching is maximal but requires~$O(\Delta)$ additional time per visited vertex. 
The running time of the $\Delta$-settling random walk is then $O( \Delta/\epsilon)$.  
\noindent We now explain how we perform edge insertions~and~deletions.

\textbf{Edge Insertion.} Our algorithm handles edge insertions as follows:
when inserting an edge $(u,v)$, if both the endpoints are free, we match it. Note that the simple algorithms stops here if at least one of the endpoints is not free.
The random walk based algorithms try to improve insertion by doing the following:
If both endpoints are matched, thus prohibiting to match the inserted edge, we do nothing.
If only one of the endpoints is matched, w.l.o.g~let this be $u$, we unmatch $u$ and $w:=\mate(u)$ and match $(u,v)$.
We then start a random walk as described above to find augmenting paths from $w$.
If the random walk is unsuccessful to further increase the size of the matching, we undo all changes and restore the matching to the state before we unmatched~$u$~and~$w$.

\textbf{Edge Deletion.}
\label{sec:rw-edge-out}
Deleting a matched edge $(u,v)$ leaves the two endpoints $u$~and~$v$ free. If possible, our algorithm matches them in $O(\Delta)$ time by scanning their neighbors in order to maintain a maximal matching. 
If $u$ and $v$ cannot be matched and the matching before edge deletion was maximal, then the matching remains maximal. However, a free vertex may be a starting point for an augmenting path of arbitrary length. Hence, we start a random walk as described above from $u$ if it is free and~do~the~same~for~$v$.

\subsubsection{Analysis}
The algorithm can maintain a $(1+\epsilon)$-approximation, if the random walks are of appropriate length and repeated sufficiently often.
More precisely, if the algorithm uses random paths of length $2/\epsilon - 1$ and the process is repeated until successful or $\Delta^{2/\epsilon-1} \log n$ times, then with high probability the matching is a $(1+\epsilon)$ approximation of the maximum matching (at each point in time). 
\begin{lemma}
\label{lemma:rw}
The random walk based algorithm maintains a $(1+\epsilon)$-approximate maximum matching if the length of the walk is $2/\epsilon-1$ and the walks are repeated $\Delta^{2/\epsilon-1} \log n$ times. \end{lemma}
\ifTR
\begin{proof}
If no augmenting path of length $\leq 2/\epsilon -1 $ exists, then the matching is a $(\frac{1/\epsilon + 1}{1/\epsilon}) = (1+\epsilon)$-approximate maximum matching. To see this, rewrite the length of the path to $2(1/\epsilon +1) -3$ and set $k=1/\epsilon+1$ in the approximation lemma above.  If there is such a path from a free node, then the probability of finding it is $\geq (\frac{1}{\Delta})^{2/\epsilon - 1}$ since one possibility is the that random walker makes the ``correct'' decision at every vertex of the path. The probability that $\lambda$ random walks of length $2/\epsilon - 1$ do not find an augmenting path of length $2/\epsilon-1$ is $\leq (1-\frac{1}{\Delta^{2/\epsilon -1}})^\lambda \leq e^{-\frac{1}{\Delta^{2/\epsilon-1}}\cdot \lambda}$. Thus for $\lambda \geq \Delta^{2/\epsilon-1} \log n$ the probability is $\leq 1/n$.
\end{proof}
\emph{Parallelization.}
Note that multiple repetitions of the random walks can be easily parallelized as they are completely independent if changes are made thread-local. If one random walker finds an augmenting path, it is accepted and the other random walkers can be stopped. \fi{}

\subsection{Blossom-based (Optimum) Algorithms}
\label{ss:optaugpath}
Note that the random walk algorithm also yields a static $(1+\epsilon)$-approximate maximum matching algorithm: use a simple greedy algorithm as initialization and then run the random walks as stated above from the remaining free nodes. However, the amount of repetitions to achieve the approximation is fairly high. Simply, running a modified BFS to find augmenting paths bounded in depth by $2/\epsilon -1$ from a free node has a theoretically faster running time $O(\Delta^{2/\epsilon-1})$ per free node. Note however that the theoretical bound for the dynamic random walk algorithm is fairly pessimistic: our algorithm stops as soon as $\emph{one}$ augmenting path has been found -- this path can also be shorter or in practice there may be multiple possibilities for augmenting paths so that the probability of finding it increases. So the natural question arises, whether a bounded augmenting path search is superior over random walk based methods stated above. Hence, we propose the following dynamic algorithms for the dynamic matching problem.

In most implementations (such as Boost \cite{2002:BGL:504206}) finding an augmenting path starting from a free node takes $\Omega(n+m)$ running time due to initialization of the data structures of the modified BFS. These data structures are initialized every time an augmenting path search is started. Hence, the observed performance of Edmonds blossom algorithm to find an optimum matching in libraries such as Boost is $\Theta(n(n+m))$ if no algorithm to initialize the matching is used and $O(F(n+m))$ if some greedy algorithm is used as initialization and $F$ is the number of remaining free nodes after greedy initialization. The later is the reason why in practice greedy initialization strategies generally help to find optimum matchings. However, finding an augmenting path can easily be implemented such that it a) stops as soon as an augmenting path is found, and b) has running time $\Theta(n' + m')$, where $n'$ and $m'$ refers to the number of nodes and edges touched by the augmenting path search modified BFS~\cite{DBLP:books/daglib/0067705,DBLP:books/cu/MehlhornN99}. The first augmenting path search needs time $O(n+m)$ to initialize the typical data structures. All searches then do book keeping of the changes they made in the data structures and undo them afterwards. Note that this clearly changes the behaviour of the algorithm in practice: if there are many short augmenting paths the algorithm will run much faster than $\Theta(n(n+m))$. The implementation does not change the worst-case complexity, but improves the best case to $O(m)$~\cite{DBLP:books/cu/MehlhornN99}. In fact, in our experience the static version of our implementation scales close to linear in $m$ in practice (as there are many short augmenting paths in real world instances). In the following, we always use this variant of augmenting path search and each of the dynamic operations does book keeping to be able to quickly search for augmenting paths.

\textbf{Edge Insertion.} Let $(u,v)$ be the inserted edge. If $u$ and $v$ are free, then we match that edge directly. Otherwise, we start an augmenting path search from $u$ if $u$ is free and from $v$ if $v$ is free.  If both $u$ and $v$ are not free, then we perform a breadth first search from $u$ to find a free node reachable via an alternating path. From this node we start an augmenting path search. Note that an augmenting path must use $(u,v)$ as both connected components did not contain an augmenting path with the component before as the algorithm maintains a maximum matching. Also note that the last case will be an expensive step in practice as the algorithm tries to maintain a maximum matching, newly inserted edges will often not result in a new augmenting path and hence the augmenting path search takes $\Theta(n+m)$ time. Without the third case of the algorithm, we call it \emph{unsafe}. That is in case both $u$ and $v$ are not free, the unsafe configuration of the algorithm does nothing.

Not using the unsafe option, the algorithm maintains a maximum matching. This is due to the fact that if the graph did not contain an augmenting path before insertion, the only way we can create one is due to the insertion of the new edge. 
The first and second case are obvious. In the third case, after finding a single free node, the augmenting path search must use the newly inserted edge $(u,v)$ (which is not matched, but both endpoints are non-free). Hence, it is sufficient to find a single free node. After running the augmenting path search, the matching size has either increased by one, or there was no augmenting path. Hence, the matching must be maximum. Lastly, note that the third case is only necessary if both endpoints of the inserted edge are in different connected components.

Note that when considering insertions only, the algorithm is more expensive than just running the static algorithm.
This is due to the fact that the static algorithm runs an augmenting path search from each free \emph{node} once, while our dynamic algorithm does try to find augmenting paths every time we insert an edge (since the graph may have changed at other places not close to the inserted edge). The overall worst case complexity in this case is $O(m(n+m))$ compared to $O(n(n+m))$ for the static algorithm. In our experiments, this effect is especially noticeable if we start a search from a node where a previous augmenting path search has been unsuccessful. 

Hence, besides using the unsafe option which drops the property that the matching is maximum, we propose the following optimization called \emph{lazy augmenting path search}. Here, we start an augmenting path search from $u$ and $v$ only if at least $m'/2$ edges have been inserted or deleted since the last augmenting path search from $u$ or $v$ or no augmenting path search has been started. Note that this effectively amortizes the cost for the augmenting path search, yielding amortized constant time per edge.
Our experiments indicate that this speeds up the overall time of the algorithm drastically, while being only slightly worse than the optimum algorithm. Our third optimization limits the search depth of the augmenting path search to $2/\epsilon-1$. This ensures that there is no augmenting path of length $2/\epsilon -1$ and hence is a deterministic $(1+\epsilon)$-approximate matching algorithm (if the deletion part algorithm ensures this as well, and the algorithm is run with the safe option). Note that the worst case complexity of the optimum version of the insertion operation is $O(n+m)$, but in practice augmenting paths (if present) are much shorter. The bounded version of our algorithm has, however, worse case complexity of $O(\Delta^{2/\epsilon-1})$.

\textbf{Edge Deletion.}
Let $(u,v)$ be the deleted edge. After the deletion we start an augmenting path search from any free endpoint $u$ or $v$. Depending on the configuration of the algorithm this either does a full run for an augmenting path or stops when the augmenting path search reached depth $2/\epsilon -1$. In the first case, this guarantees that the matching is maximum if it was maximum before and in the latter case, our algorithm maintains an $(1+\epsilon)$ approximate maximum matching. 
If case we use lazy augmenting path search,
we start an (depth bounded) augmenting path search from $u$ and $v$ only if at least $m'/2$ edges have been inserted or deleted since the last augmenting path search from $u$ or $v$.
Otherwise, we limit augmenting path search from $u$ and $v$ to augmenting paths of length $\min(3, 2/\epsilon -1)$.

\subsection{Baswana, Gupta and Sen Algorithm}
\label{sec:bgs}

Baswana, Gupta and Sen (BGS) presented an randomized algorithm in \cite{BaswanaGS15}, that maintains a maximal matching in a dynamic graph in \emph{amortized} $O(\sqrt{n})$ update time with high probability. They also present a multi-level  variant that runs in $O(\log n)$ amortized time. To be self contained, we briefly review the main concepts of the algorithm and follow their description closely.
\csch{TODO: check two level vs log n levels + theory behind it}

\textbf{Levels and Ownership of Edges.}
\label{sec:own-edge}
\ifTR
\begin{algorithm}[b!]
	Let $(u,v)$ be a uniformly randomly selected edge from $\mathcal{O}_u$\;

	\ForAll{$(v,w) \in \mathcal{O}_u$}{
 		$\text{remove } (v,w) \text{ from } \mathcal{O}_w$\;
	}
 	\If{$v$ is matched}{
                $x \gets$ mate($v$); 
		$M \gets M \setminus \{(v,x)\}$\;
 	} \Else {
                $x \gets$ NULL
        }
 	$M \gets M \cup \{(u,v)\}$, $\textsc{level}(u) \gets 1,\,\textsc{level}(v) \gets 1$\;

 	\Return{$z$}\;
\caption{\textsc{Random-Settle}($u$): find a random edge $(u,v)$ from the set of owned edges of $u$, matches it and returns the previous mate of~$v$.}
\label{pscd:bgs-rs}
\end{algorithm} \fi{}
The algorithm uses the concept of \emph{ownership} for edges.
More precisely, based upon the number of edges that a vertex \emph{owns}, the algorithm partitions the set of vertices into two levels 0 and 1. 
An edge is always owned by at least one of its endpoints. 
If both endpoints are at level 0, then both vertices own the edge. 
If only one endpoint is at level 1, then this endpoint owns the edge. 
If both endpoints are at level 1, then exactly one endpoint, namely the first mentioned vertex owns the edge. If a new edge $(u,v)$ with $\mathit{level}(u) = \mathit{level}(v) = 1$ is inserted, it will therefore be owned by the vertex $u$. The set $\mathcal{O}_u$ denotes the set of edges owned by a vertex $u$. The level of an edge is defined by $\mathit{level}(e = \{u,v\})=\max(\mathit{level}(u),\mathit{level}(v))$.
 BGS maintains the following invariants: 
(1) Every vertex on level 1 is matched.
(2) Every free vertex on level 0 has all neighbours matched.
(3) Every vertex on level 0 owns less then $\sqrt{n}$ edges (at any moment of time).
(4) Both endpoints of each matched edge are on same level.

\textbf{Edge Insertion.}
\label{sec:bgs-edge-in}
Let $(u,v)$ be the edge being inserted.
If either $u$ or $v$ are at level~$1$, then there is no violation of any invariant.
The algorithm adds $(u,v)$ to~$\mathcal{O}_u$ if level$(u) = 1$ and to $\mathcal{O}_v$ otherwise.
If both endpoints of $(u,v)$ are at level~0, then the algorithm proceeds as follows: 
If both endpoints are free, the edge is added to the matching. 
Adding the edge $(u,v)$ to the sets $\mathcal{O}_u$ and $\mathcal{O}_v$ increases the number of edges owned by $u$ and $v$. 
If at least one set $\mathcal{O}_u$ or $\mathcal{O}_v$ exceeds the threshold of $\sqrt{n}$ in size, the vertex with the higher number of owned edges will be \emph{repaired}. 
Let $u$ be that vertex.
Repairing a vertex $u$ is done by calling the procedure \textsc{Random-Settle} on $u$. As a result, $u$ moves to level 1 and gets matched to some vertex $y$ selected randomly uniformly from the set of owned edges $\mathcal{O}_u$. 
The vertex $y$ is also moved to level 1 to satisfy invariant 4.
If $w$ and $x$ were the earlier mates of $u$ and $y$ at level 0, respectively, then matching $u$ and $y$ has rendered $w$ and $x$ free. 
The algorithm tries to settle each of those by scanning their set of owned edges for free vertices.

\textbf{Edge Deletion.}
\label{sec:bgs-edge-out}
Let $(u,v)$ be an edge that is deleted.
If the edge has not been matched, then after removing the edge from the graph all invariants still hold.
If it has been matched, then $u$ and $v$ are now free.
Therefore, the first invariant may be violated.
If $(u,v)$ is at level 0, then the algorithm tries to settle both endpoints by scanning their sets of owned edges.
If $(u,v)$ is at level 1, then $u$ the algorithm does the following:
First, $u$ disowns all its edges whose other endpoint is at level 1. If $\mathcal{O}_u$ is still greater than or equal to $\sqrt{n}$, then $u$ stays at level 1 and executes \textsc{Random-Settle}($u$). If
$u$ owns less than $\sqrt{n}$ edges, it moves to level 0 and tries to settle it by scanning its set of owned edges. 
The transition of $u$ from level 1 to 0 leads to an increase in the number of edges owned by each of its neighbors at level 0. 
This may violate the size constraint of owned edges for those neighbors.
Hence, the algorithm calls \textsc{Random-Settle} for each neighbor that violates the~constraint, which moves it to level 1.

\subsection{Neiman and Solomon Algorithm}
\label{sec:ns}

In contrast to the BGS algorithm \cite{BaswanaGS15}, which is randomized, Neiman and Solomon (NS) \cite{NeimanS16} present a deterministic algorithm for maintaining a maximal matching in a dynamic graph. Their approach guarantees, that the maintained matching is a $3/2$-approximate maximum matching and that update time is $O(\sqrt{m})$ in \emph{worst case}, where $m$ denotes the number of edges present in the graph in the moment of the update.
NS maintains the following invariants: 
There are no augmenting paths of length $\leq 3$, ensuring $3/2$-approx matching.
All free vertices have degree at most $\sqrt{2n + 2m}$.

\begin{lemma}{\cite{NeimanS16}}
\label{lemma:ns-sqrt-m}
Any free vertex of degree larger than $\sqrt{m}$ can always be matched, so as to generate a free vertex with degree less than $\sqrt{m}$. This can be achieved in $O(\sqrt{m})$ time.\csch{where is this from?, do we need the proof?}
\end{lemma}
\ifTR
\begin{proof}
For a high degree vertices $u$, one can find a surrogate $z$ with degree of at most $\sqrt{2m}$ which is the mate of a neighbour of $u$. To find such a surrogate, one scans linear through the neighbours of $u$, retrieving $z=\mate(w), w \in N(u)$, where $N(u)$ denotes the set of neighbours of $u$, and then checking if $\deg(z)\leq\sqrt{2m}$. Clearly, if $w$ is free one can simply match $u$ and $w$, else among $O(\sqrt{m})$ neighbors of $u$ one finds a low degree surrogate $z$, as otherwise there would be $(\sqrt{m}+1)\cdot\sqrt{m}=m+\sqrt{m} > m$ edges in the graph, which is not possible. Then unmatch the edge $(w,z)$ and  match $(u,w)$ to ensure the created free vertex $z$  has $\deg{z}<\sqrt{2m}$.
\end{proof}\fi{}

\textbf{Edge Insertion.}
\label{sec:ns-edge-in}
Let $(u,v)$ be the edge being inserted. If both the endpoints are free, the edge is simply added to the matching. Also, if both endpoints are matched it does not entail any further processing. However, if exactly one endpoint of the edge, say $u$, is matched, they try to remove a possible augmenting path of length $3$ as follows.
The neighbours of the mate of $u$ say $u'=\mate(u)$, are scanned for a free vertex, say $x$. If such a free vertex exists, an augmenting path of length $3$ has been found, which is augmented increasing the matching~size. 
 
\textbf{Edge Deletion.}
\label{sec:ns-edge-out}
Let $(u,v)$ be the edge being deleted. If the edge was unmatched, its deletion cannot create any new augmenting paths. However, if it was a matched edge, both the endpoints become free after the edge deletion. First, the algorithm checks for both freed vertices whether they have free neighbours and if so matches the freed vertices with those free neighbours. Now, in order to eliminate augmenting paths of length $3$ starting from a free vertex, say $u$, all neighbours $w$ of $u$ are scanned checking if $\mate(w)$ has a free neighbour. By providing appropriate data structures, this can be done in $O(\sqrt{n})$ time. If an augmenting path has been found, it is augmented increasing the size of the matching by one. 
If no augmenting path has been found, vertex $u$ remains free, but only if its degree is at most $\sqrt{2m}$. If the degree of $u$ exceeds $\sqrt{2m}$, using Lemma \ref{lemma:ns-sqrt-m} a surrogate can be found in $O(\sqrt{m})$. 

\noindent
The overall update time of the algorithm is bound by the bounded degree of all free vertices, making any linear search through the neighbourhood $N(u)$ of a vertex $u$ cost at most $O(\sqrt{n+m})$. Bounding the degree can further be achieved in $O(\sqrt{m})$ time using Lemma \ref{lemma:ns-sqrt-m}.

\section{Experimental Evaluation}\label{sec:eval}
\textbf{Implementation and System.}\label{Methodology}
We implemented the algorithms described in the previous section.
The codes are written written in C++ and have been compiled using g++-7.3.0 with flags \texttt{-O3}. All codes are sequential.
We plan to further improve the codes and then to release them to make it available to a larger audience.
Our experiments are conducted on one core of a
machine with AMD Opteron Processors 6174 with 2.2GHz and 256GB of RAM.
\ifTR\emph{Dynamic Graph Data Structure:} our algorithms use the following dynamic graph data structure. For each node $v$, we maintain a vector $L_v$ of adjacent nodes, and a hash table $\mathcal{H}_v$ that maps a vertex $u$ that is incident to $v$ to its position in $L_v$. This data structure allows for expected constant time insertion and deletion as well as a constant time operation to select a random neighbor of $v$. The deletion operation on $(v,u)$ is implemented as follows: get the position of $u$ in $L_v$ via a lookup in $\mathcal{H}_v(u)$. Swap the element in $L_v$ with the last element $w$ in the vector and update the position of $w$ in $\mathcal{H}_v$. Finally, pop the last element (now $u$) from $L_v$ and delete its entry from $\mathcal{H}_v$.\fi{}

\noindent \textbf{Instances and Methodology.}
By default we perform ten repetitions per instance.
We measure the total time taken to compute all edge insertions and deletions
and generally use the \emph{geometric mean} when averaging over different instances
in order to give every instance a comparable influence on the final result. 
In order to compare different algorithms, we use \emph{performance profiles}~\cite{DBLP:journals/mp/DolanM02}.
These plots relate the matching size / running time  of all algorithms to the corresponding matching size / running time produced / consumed by each algorithm.
More precisely, the $y$-axis shows $\#\{\text{objective} \geq \tau * \text{best} \} / \# \text{instances}$, where objective corresponds to
the result of an algorithm on an instance and best refers to the best result of any algorithm shown within the plot.
When we look at running time, the $y$-axis shows $\#\{$t$ \leq \text{fastest}/\tau \} / \# \text{instances}$, where $t$ corresponds to
the time of an algorithm on an instance and fastest refers to the time of the fastest algorithm on that instance.
The parameter $\tau\leq 1$ in this equation is plotted on the $x$-axis.
For each algorithm, this yields a non-decreasing, piecewise constant function.
Thus, if we are interested in the number of instances where an algorithm is the best/fastest, we only need to look at $\tau = 1$.

\noindent \textbf{\textit{Instances.}} We evaluate our algorithms on a number of large graphs. 
These graphs are collected from
      \cite{benchmarksfornetworksanalysis,UFsparsematrixcollection,snap,DBLP:conf/www/Kunegis13,konect:unlink}.
      Table~\ref{tab:graphstable} summarizes the main properties of the benchmark set.
      Our benchmark set includes a number of graphs from numeric simulations as well as complex networks.
These include static graphs as well as real dynamic graphs.
As our algorithms do only handle undirected graphs, we consider all input graphs to be undirected by ignoring edge directions and we remove self-loops and parallel edges.
We perform \emph{two different types} of experiments. 
First, we use the algorithms using insertions only, \ie we start with an empty graph and insert all edges of the static graph in a random order. We do this with all graphs from~Table~\ref{staticgraphs}.
Second, we use real dynamic instances from Table~\ref{dyninstances}. Most of these instances, however, only feature insertions (with the exception being \texttt{dewiki} and \texttt{wiki-simple-en}). 
Hence, we perform additional experiments with fully dynamic graphs from these inputs, by undoing $x$ percent of the update operations performed last.



\subsection{Random Walk and Blossom-based Algorithms}
In this section, we use our algorithms with random insertions only.
More precisely, we use the static graphs from Table~\ref{staticgraphs}. For each experiment, we start with an empty graph and insert edges of the static input in random order until all edges are inserteall edges are insertedn compare the result of our dynamic algorithms the maximum matching on the final graph~Edmond~\cite{edmonds1965paths}. 

\textbf{\emph{Random Walk-Based Algorithms:}} We start with random walk-based algorithms. Preliminary experiments have shown that decreasing $\epsilon$ is more effective in getting better solutions than performing more repeated random walks at the start node. Hence, we exclude algorithms that perform multiple repetitions of random walks per insert operation here from the evaluation and focus on the different values of $\epsilon$. We vary $\epsilon \in \{0.1, 0.25, 0.5, 1,2\}$. Recall that the path length of a single random walk is then bounded by $2/\epsilon-1$. If \emph{all} paths of that length were explored, the algorithms would be guaranteed to give a $(1+\epsilon)$-approximation. Figure~\ref{fig:rwperfprofile} summarizes the result. It is not surprising that the algorithm needs more running time for smaller $\epsilon$, but also yields better results (see Figure~\ref{fig:rwperfprofile}) with increasing path lengths. 
On average, the algorithm is 2.4\%, 3.2\%, 4.2\%, 5.5\%, 11.5\% percent away from the optimum for $\epsilon=0.1, 0.25, 0.5, 1, 2$ respectively.
Thus, even though the algorithms are not guaranteed to explore all paths of length $2/\epsilon- 1$, they achieve in practice an approximation that is much better than the theoretical bound for algorithms that explore all such paths.
\ifTR
\begin{figure}[b!]
\else
\begin{figure}[t!]
\fi{}
\centering
\vspace*{-.75cm}
\ifTR
\includegraphics[width=6.5cm]{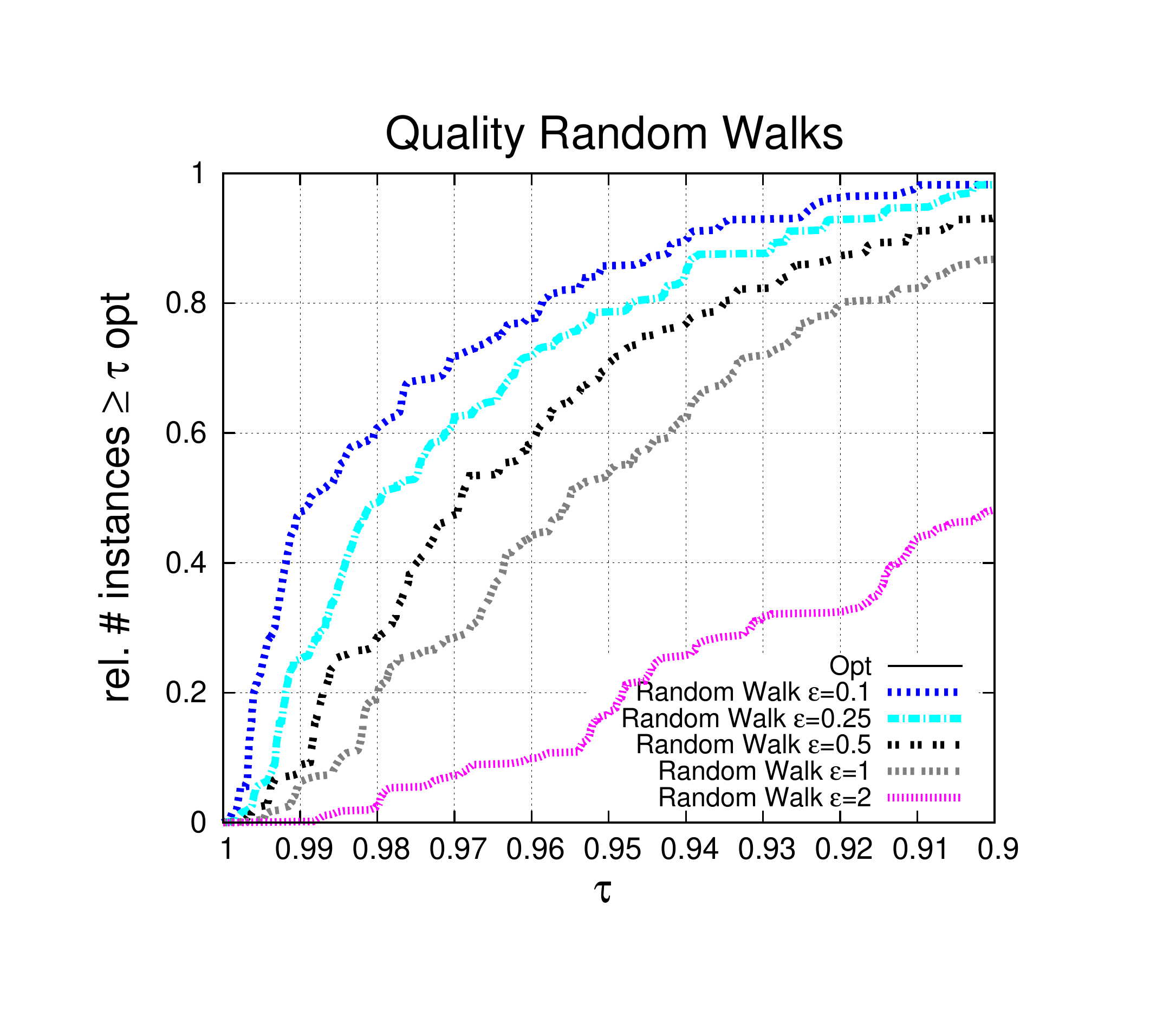}
\includegraphics[width=6.5cm]{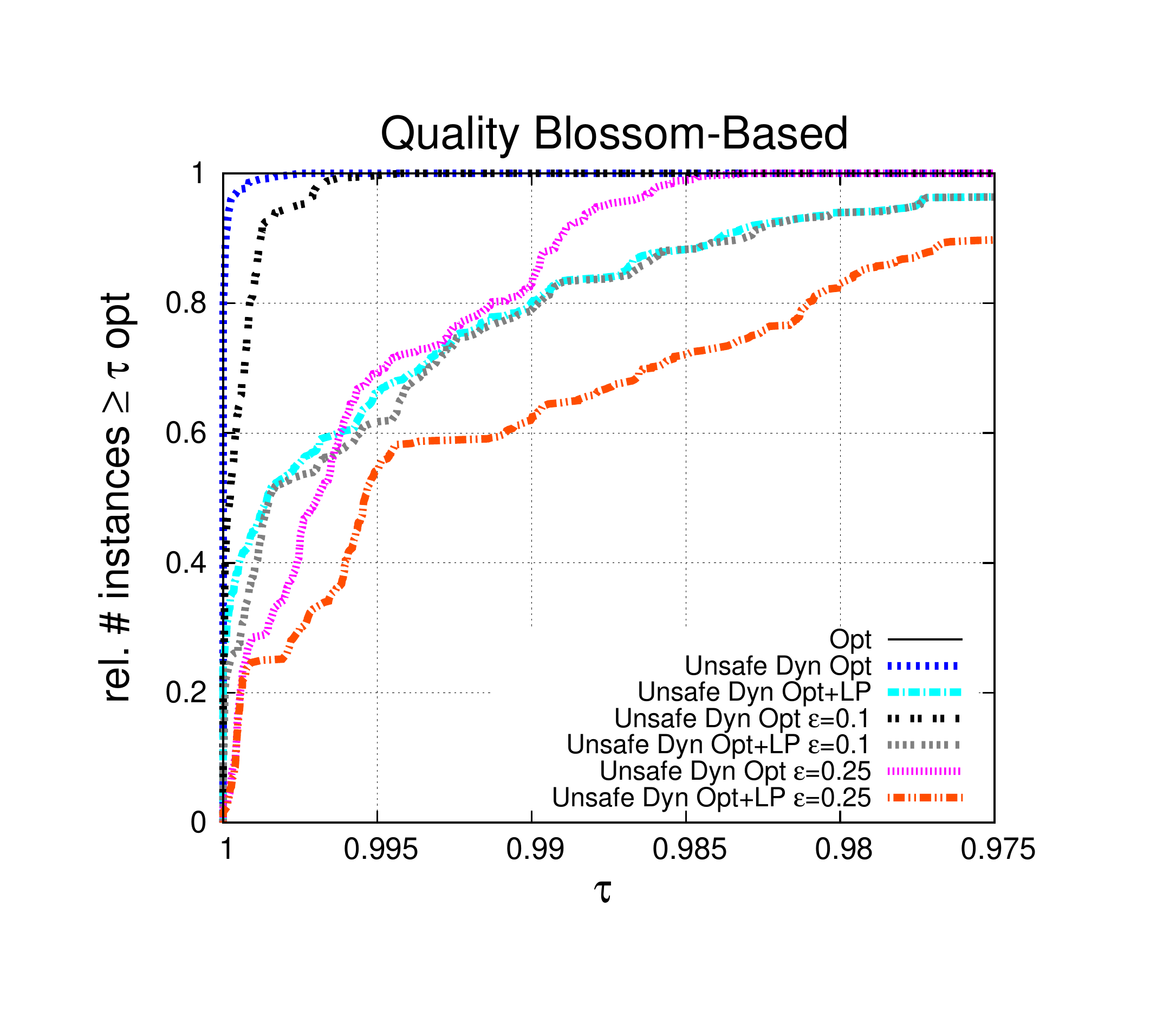}
\fi{}
\vspace*{-.5cm}
\caption{Performance profile for matching size $|\mathcal{M}|$ for Random Walk and for Unsafe Dyn Opt. }
\vspace*{-.5cm}
\label{fig:rwperfprofile}
\end{figure}
The strongest configuration ($\epsilon=0.1$) is at most 1\% away from the optimum matching size in 50\% of the cases (see Figure~\ref{fig:rwperfprofile}). 
Note that the random walk algorithm does not achieve the guarantee of 1\% approximation as claimed by Lemma~\ref{lemma:rw} since we did not perform the vast amount of repetitions necessary to get the result in expectation -- instead we performed a single repetition of the random walker for each insertion. 
As excepted running time does increase with decreasing $\epsilon$. However, due to random walks that can finish early because they managed to match an edge, the effect is less visible than theory expects. The running time increase over the random walk using $\epsilon=2$ (which is essentially a random walker not allowed to move, and hence boils down to the very simple greedy algorithm), is 12\%, 17\%, 21\%, 27\% for $\epsilon=1, 0.5, 0.25, 0.1$, respectively.

Enabling $\Delta$-settling generally improves the result.
On average, the random walk with $\Delta$-settling is now 1.1\%, 1.4\%, 1.8\%, 2.2\%, 3.7\% away from the optimum matching for $\epsilon=0.1, 0.25, 0.5, 1, 2$, respectively.
On average in our experiments using $\Delta$-settling has a negligible impact on running time.  Hence, we recommend to use $\Delta$-settling when using random walk-based algorithms and do so in the following unless otherwise mentioned.

\textbf{\emph{(Optimum) Blossom-Based Algorithms}:} We now consider dynamic blossom-based algorithms from Section~\ref{ss:optaugpath}. We start this section with the version of the algorithm that maintains the optimum matching, and compare it to the naive dynamic optimum matching algorithm that recomputes a maximum matching from scratch each time an edge is inserted. Since the running time of the naive optimum algorithm is fairly excessive, we run it only on the graphs of our benchmark set having less than 25k nodes. First of all, our dynamic algorithm that maintains the optimum matching is more than an order of magnitude faster than the naive optimum algorithm (roughly a factor 12). We expect that the difference will be even more pronounced if even larger graphs are used. Running our dynamic optimum algorithm with the unsafe option indeed significantly speeds up the algorithm -- the lazy augmenting path search configuration is more than two orders of magnitude over the safe version of our algorithm (roughly a factor 115). 
The improvements in running time stem from the fact that our algorithms try to maintain a very large matching. Hence, the case that is executed by the safe option often does not find an augmenting path which implies that the augmenting path search has to look at the overall network and hence reaches its worst-case complexity.
Of course, the unsafe option does not have a guarantee on optimality anymore. In our experiments, the unsafe option computes matchings that are 0,02\% worse than the optimum on average. We conclude that the algorithm maintains near-optimum matchings while being three orders of magnitude faster than the naive optimum dynamic algorithm.  Henceforth, we only consider the unsafe version of our algorithm.

We now switch our set of graphs back to all of our benchmark graphs~from Table~\ref{staticgraphs}. Using lazy augmenting path search in the unsafe algorithm additionally speedups up computations. Unsafe+LP is on average 20.5 faster than the unsafe algorithm without lazy augmenting path search -- again at the cost of solution quality. The algorithm is already only 30\% slower than running the static algorithm a \emph{single} time on the final graph that contains all edges. On the other hand, the unsafe dynamic algorithm using lazy augmenting path search computes 0.6\% worse matchings that the unsafe algorithm without lazy augmenting path search.

Lastly, we focus on the third variation of the algorithm, which is to bound the depth of the augmenting path search to that is done during update operations. The depth is bounded to $2/\epsilon-1$ so that given $\epsilon$ and running the safe option of the algorithm would maintain a $1+\epsilon$ approximate matching. We, however, only consider the unsafe version of the algorithm. We use same values of $\epsilon=0.1, 0.25, 0.5, 1$ as in the random walk-based algorithms section, but do not consider $\epsilon=2$, since this is again essentially the very simple greedy algorithm. Moreover, we run the algorithm with and without the lazy augmenting path search.
\begin{figure}[t!]
\centering
\vspace*{-.5cm}
\ifTR
\includegraphics[width=6.5cm]{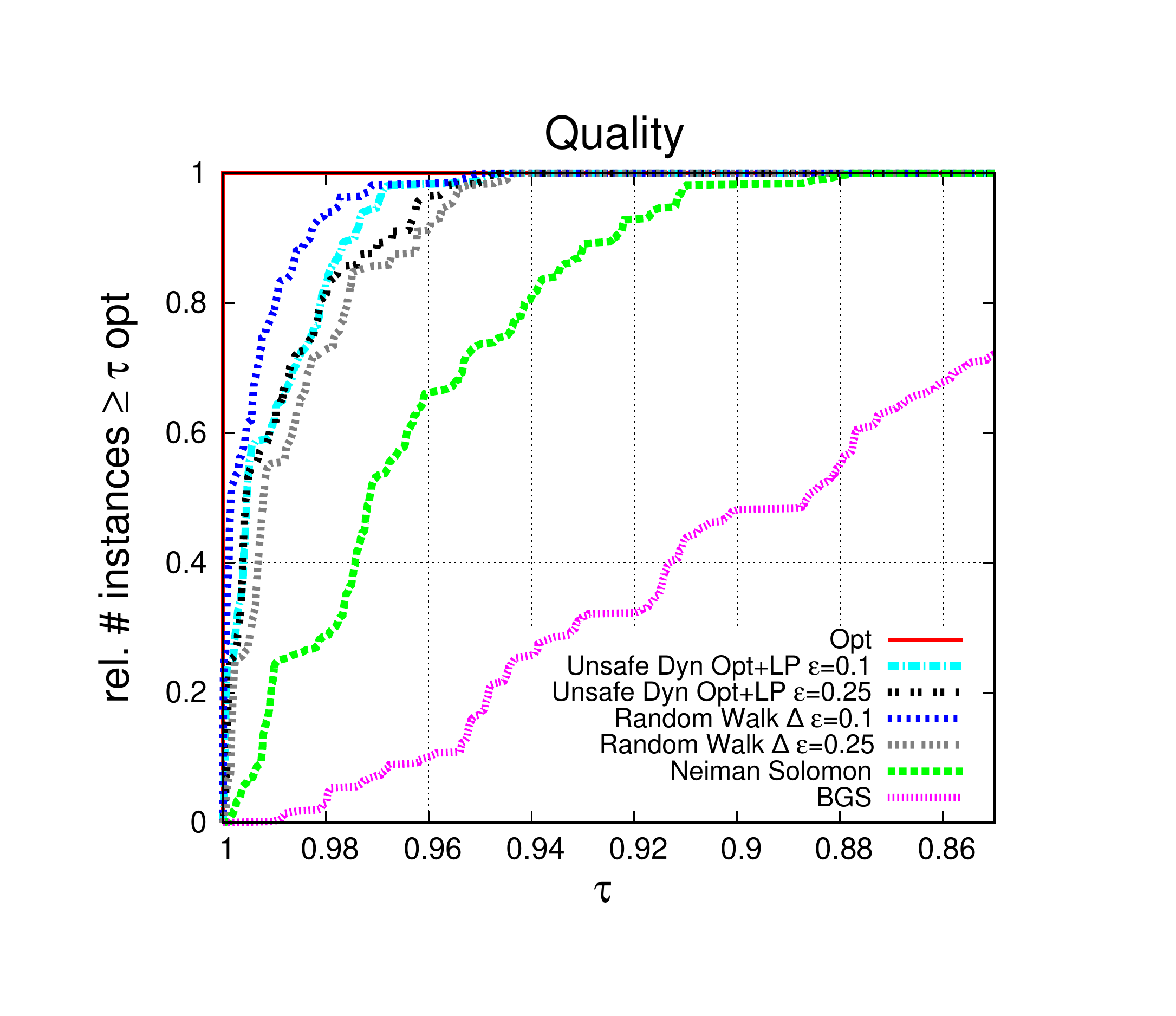}
\includegraphics[width=6.5cm]{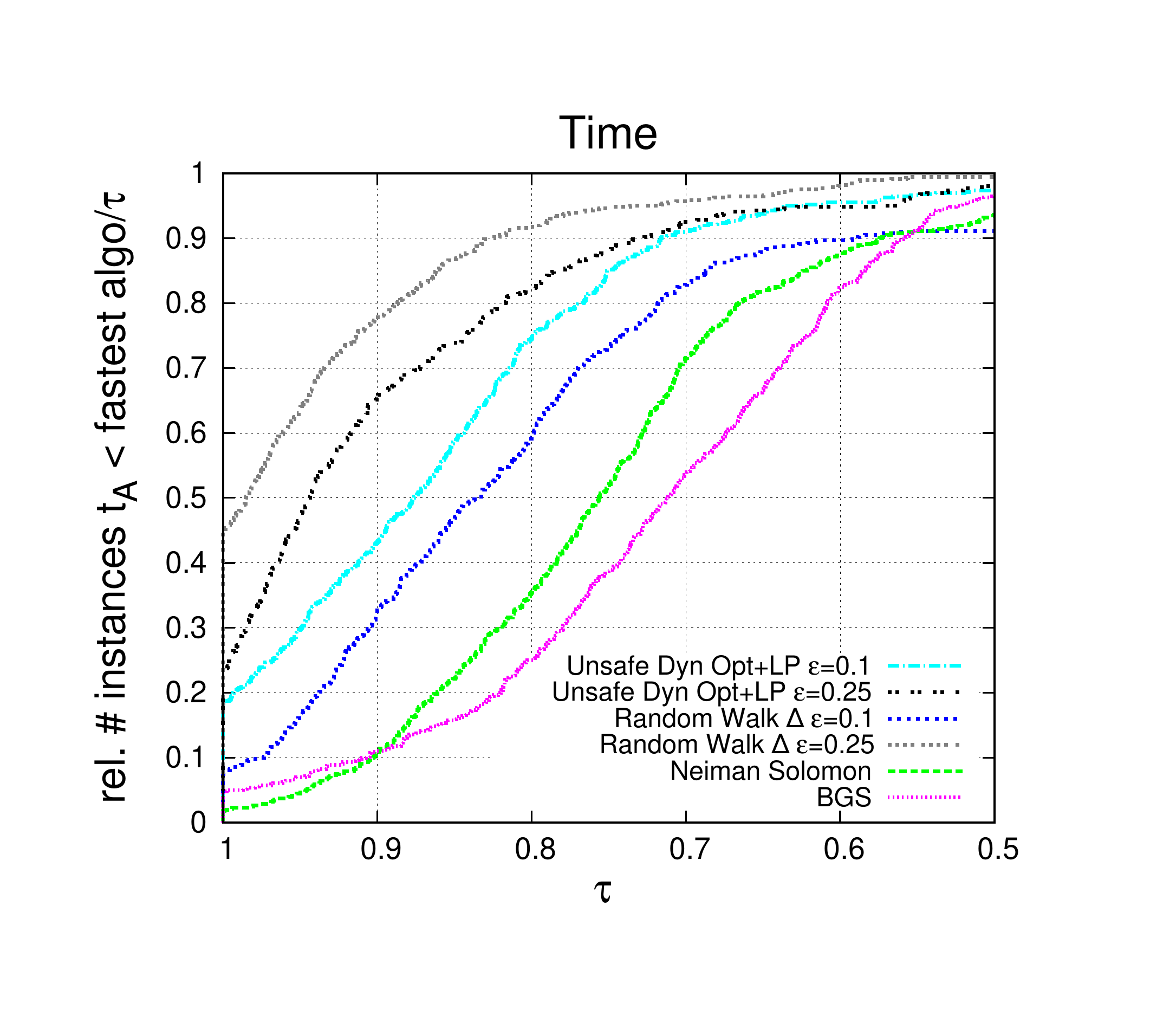}
\fi{}
\vspace*{-.5cm}
\caption{Performance profile for matching size $|\mathcal{M}|$ and time for all algorithms. In all cases, if an algorithm has a curve closer to the upper left corner, then the algorithm is better.}
\vspace*{-.5cm}
\label{fig:allperfprofile}
\end{figure}

First of all, running without lazy augmenting path search, the algorithm indeed maintains the approximation guarantee. On average, the algorithm is 0.1\%, 0.4\%, 1.5\% and 3.6\% worse than the optimum algorithm for $\epsilon=0.1, 0.25, 0.5, 1$, respectively. Using the lazy augmenting path search, speeds up to algorithm by a factor of 5.52, 2.65, 2.03, 1.76 for $\epsilon=0.1, 0.25, 0.5, 1$, respectively. With lazy augmenting path search, the algorithm is 0.6\%, 1.00\%, 2.2\%, 4.3\% worse than the optimum for $\epsilon=0.1, 0.25, 0.5, 1$, respectively (and hence still achieves the approximation guarantee). The algorithm using $\epsilon=0.1$ is only 0.2\% worse than the algorithm not bounding the depth. However, the algorithm is also not much faster. On average, bounding the search depth with $\epsilon=0.1$ improves running time by 6\%. Figure~\ref{fig:rwperfprofile} shows a summarizing performance profile.

\subsection{Comparison of Algorithms }

\textbf{Dynamic Sequences from Static Graphs:} We now compare all of the different non-optimal algorithms against each other for the insertion-only case. 
For random-walks, we always enable $\Delta$-settling, for blossom-based algorithm always use the unsafe option and with and without lazy augmenting path search.
Table~\ref{tab:averageresults} shows average results for matching size and running time after all edges and operations have been performed.
Figure~\ref{fig:allperfprofile} shows performance profiles for running time and for matching size. 

First of all, both the blossom-based (with lazy augmenting path search) and random walk-based algorithms dominate the algorithms by Neiman Solomon and Baswana Gupta Sen (BGS). The algorithms find consistently larger matchings and do so in less time. However, note that the real-world instances we look at never have more than $\sqrt{n}$ edges, so that the BGS algorithm is roughly similar to the simple greedy algorithm. We also try to use $c*\sqrt{n}$ as a threshold for different values of $c$, but this always resulted in worse matching sizes.
\ifTR
\begin{table}[b!]
\else
\begin{table}[t!]
\fi{}
\centering
\caption{Random insertions from static graphs: mean of the matching size relative to optimum  after all operations have been done as well as mean increase in running time over Random Walk,$\Delta$,$\epsilon=0.25$.}
\label{tab:averageresults}
{}
\begin{tabular}{llrr}
\toprule
algorithm & & mean $|\mathcal{M}|$ / $|\mathcal{M}_\text{opt}|$  &rel. time \\
                  \midrule
BGS                  &                 &0.885 & 32\% \\
Neiman Solomon       &                 &0.964 & 28\%\\
\textbf{Unsafe Dyn Opt+LP}    & $\epsilon=0.1$  &0.994 & 27\%\\
\textbf{Unsafe Dyn Opt+LP}    & $\epsilon=0.25$ &0.990 & 11\%\\
Unsafe Dyn Opt       & $\epsilon=0.1$  &0.999 & 613\%\\
Unsafe Dyn Opt       & $\epsilon=0.25$ &0.996 & 192\%\\
Unsafe Dyn Opt       & $\epsilon=0.5$  &0.985 & 101\%\\
Unsafe Dyn Opt       & $\epsilon=1$    &0.964 & 67\%\\
\textbf{Random Walk,$\Delta$} & $\epsilon=0.1$  &0.989 & 5\% \\
\textbf{Random Walk,$\Delta$} & $\epsilon=0.25$ &0.986 & 1\\
         \bottomrule
\end{tabular}
\vspace*{-.5cm}
\end{table} 

In general, performance differences in running time are not very big (except if we don't use lazy augmenting path search in the blossom-based algorithms). 
Secondly, for the same values of $\epsilon$ the blossom-based algorithms compute slightly better results than their random walk-based counter parts. This is not surprising as the blossom-based algorithms explore larger subgraphs for each edge that has been inserted. We conclude here that both types of algorithms are feasible in practice and have an advantage in solution quality over Neiman Solomon and Baswana Gupta Sen on graphs with random insertions. Moreover, both of these algorithm yield a clear trade-off between running time and solution quality via the $\epsilon$ parameter. On the other hand, all of the algorithms considered here are roughly five orders of magnitude faster than the naive dynamic optimum algorithm (only considering instances having less than 25k nodes) and except Baswana Gupta Sen, all of these algorithms are within a 4\% range of the optimum matching size.

\noindent\textbf{Real-World Dynamic Instances:}
\label{sec:realdyn}
We now switch to the real-world dynamic instances. As already mentioned, most of these instances are insertion-only.
Hence, we perform additional experiments with fully dynamic graphs from these inputs, by undoing $x$ percent of the update operations performed last (call them  $\mathcal{O}_x$). More precisely, we perform the operations in $\mathcal{O}_x$ in reverse order. More precisely, if an edge operation was an insertion in $\mathcal{O}_x$, we perform a delete operation and if it was a delete operation we insert it.
As before, we compute the update on the graph after each removal/insertion. The connection to practice in this case, is that with undoing operations, we want to restore a previous state. Table~\ref{tab:averageresultsdyn} summarizes the results of the experiment and Figure~\ref{fig:real} compares the algorithms on the two real-world dynamic graphs dewiki and wiki\_simple\_en.
\begin{table}[t!]
\centering
\caption{Real-world dynamic instances: mean of the matching size relative to optimum after all operations have been done as well as the mean increase in running time over Random Walk,$\Delta$,$\epsilon=0.5$.}
\label{tab:averageresultsdyn}
{}
\begin{tabular}{llrrrr}
\toprule
\# undo op & & 0& 5\% & 10\% & 25\% \\
                \midrule
algorithm & & \multicolumn{4}{c}{mean $|\mathcal{M}|$ / $|\mathcal{M}_\text{opt}|$}   \\
                  \midrule
BGS                 &                &0.845  &0.847&0.848 &0.851 \\
Neiman Solomon      &                &0.968  &0.971&0.973 &0.976\\
Unsafe Dyn Opt+LP   & $\epsilon=0.1$  &0.947  &0.985&0.990 &0.996\\
Unsafe Dyn Opt+LP   & $\epsilon=0.25$ &0.942  &0.982&0.988 &0.993\\
Unsafe Dyn Opt      & $\epsilon=0.\overline{33}$ &0.994  &0.996&0.997 &0.998\\
Unsafe Dyn Opt      & $\epsilon=0.5$  &0.988  &0.991&0.992 &0.994\\
Unsafe Dyn Opt      & $\epsilon=1$    &0.968  &0.971&0.973 &0.976\\
\textbf{Random Walk,$\Delta$} & $\epsilon=0.1$  &0.982  &0.984&0.985 &0.986\\
\textbf{Random Walk,$\Delta$} & $\epsilon=0.25$ &0.981  &0.983&0.984 &0.985\\
\textbf{Random Walk,$\Delta$} & $\epsilon=0.5$  &0.978  &0.980&0.981 &0.982\\
       \midrule
algorithm &  & \multicolumn{4}{c}{rel. time}   \\
       \midrule
BGS                  &                            & 4\%    & 14\%   & 13\%&18\%\\
Neiman Solomon       &                            & 64\%   & 82\%   & 92\%&112\% \\
Unsafe Dyn Opt+LP    & $\epsilon=0.1$             & 82\%   & 306\%  & 383\%& 633\%\\
Unsafe Dyn Opt+LP    & $\epsilon=0.25$            & 23\%   & 149\%  & 200\% & 346\%\\
Unsafe Dyn Opt       & $\epsilon=0.\overline{33}$ & 1\ 551\% & 1\ 551\% &1\ 598\% & 1814\%\\
Unsafe Dyn Opt       & $\epsilon=0.5$             & 679\%  & 682\%  &713\% & 800\%\\
Unsafe Dyn Opt       & $\epsilon=1$               & 210\%  & 212\%  & 223\% & 250\%\\
\textbf{Random Walk,$\Delta$} & $\epsilon=0.1$             & 25\%   & 26\%  & 24\% & 24\%\\
\textbf{Random Walk,$\Delta$} & $\epsilon=0.25$            & 10\%   & 11\%  & 9\% & 11\%\\
\textbf{Random Walk,$\Delta$} & $\epsilon=0.5$             & 1       & 1      & 1 & 1\\

\bottomrule
\vspace*{-.5cm}
\end{tabular}
\end{table} 
{}

Overall, the situation is similar to experiments with random insertions that we have seen before.
The random walk with $\Delta$-settling and $\epsilon=0.5$ dominates Baswana, Gupta, Sen and Neiman Solomon in terms of running time \emph{and} matching size for every number of undo operations. The blossom-based algorithm with lazy path search, however, yields smaller matchings if no operations are undone. We believe that this is due to the edges not being inserted randomly and hence the lazy augmenting path search heuristic is less effective, and misses augmenting paths that have been created over time. If operations are undone, the blossom-based algorithms outperform Neiman Solomon in terms of matching size, but are also considerably slower as the deletion operations search for augmenting paths of lengths three (except for $\epsilon=1$). The blossom-based algorithm without lazy augmenting path search get very close to the optimum solutions. The best algorithm here is blossom-based algorithm without lazy augmenting path search for $\epsilon=1/3$. On average, it computes solutions that are $<0.6\%$ away from the optimum (for every amount of undo operations done).

In general, all algorithms improve quality relative to the optimum matching size, if we undo operations. This is due to the fact that the matching may have changed over time and hence new (short) augmenting paths may be found. In case of random-walks this is also simply due to the fact that additional work is performed and the likelihood to find an augmenting path is increased by running additional random walks. 
Summing up, all of the algorithms, except Baswana, Gupta, Sen, compute/maintain very large matchings.
Blossom-based and random walk-based algorithm are highly flexible and are able to trade solution quality for time. Overall, random walk-based algorithms seem to be the method of choice in practice. 
\vspace*{-.25cm}
\section{Conclusion}
\ifTR Few theoretical
algorithmic ideas of fully dynamic maximal matching algorithms have been tried out in practice. 
We started to bridge the gap between theory and practice that is currently observed for the problem. \fi{}
We looked at several dynamic matching algorithms including Baswana, Gupta and Sen \cite{BaswanaGS15}, Neiman and Solomon~\cite{NeimanS16}, as well as random walk-based algorithms and blossom-based algorithms.
We performed extensive experiments comparing the performance of these algorithms on the real-world and artificially generated instances. 
In terms of results, first we have shown that maintaining optimum matchings can be done much more efficiently than the naive algorithm that recomputes maximum matchings from scratch. Second, we have seen that all non-optimum dynamic algorithms that we considered in this work are able to maintain near-optimum matchings in practice while being multiple orders of magnitudes faster than the naive optimum dynamic algorithm. In practice, random walk-based algorithms with $\Delta$-settling will be the method of choice.

\ifTR
\begin{figure}[t!]
\vspace*{-1cm}
\centering
\includegraphics[width=6.5cm]{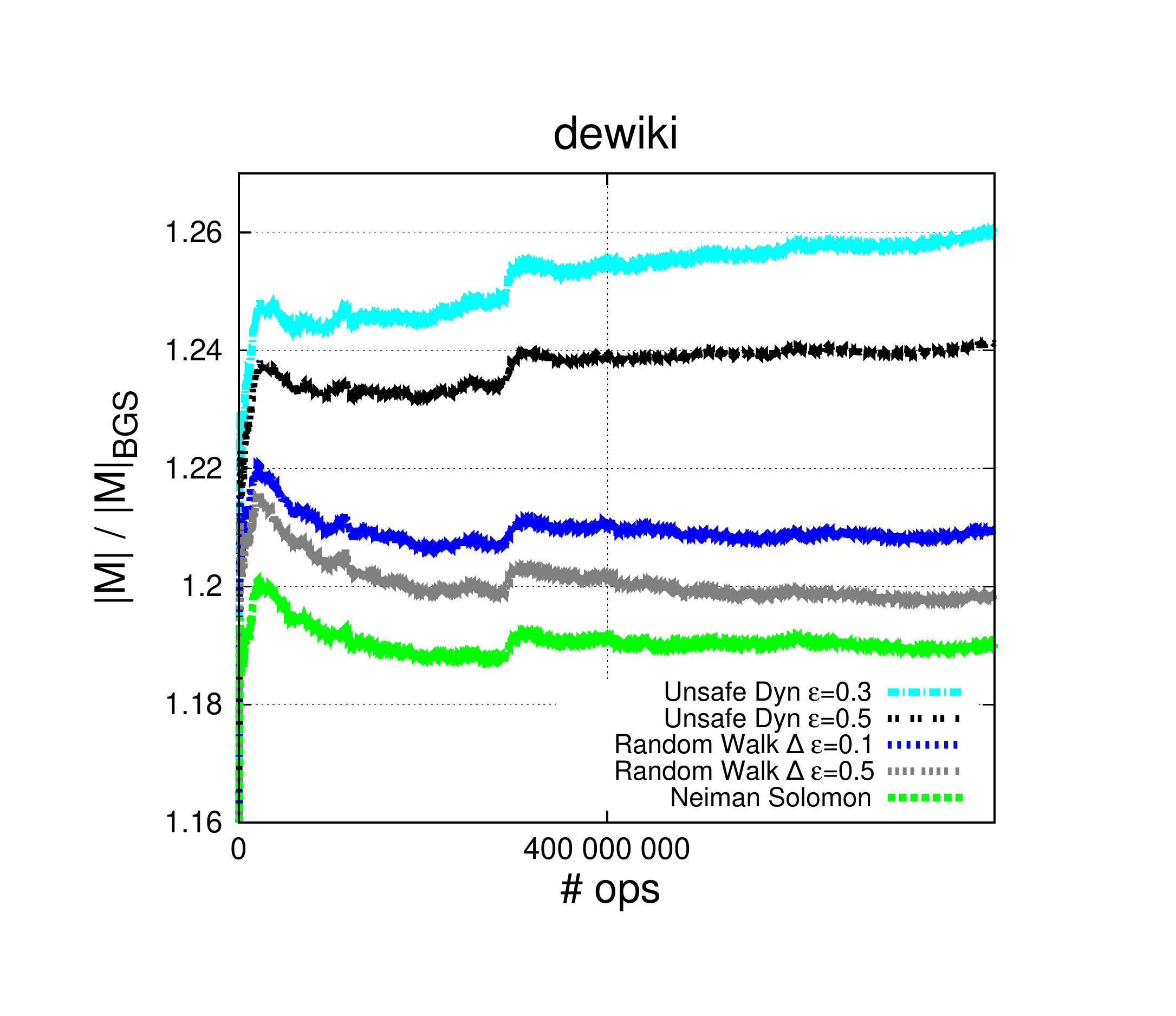}
\includegraphics[width=6.5cm]{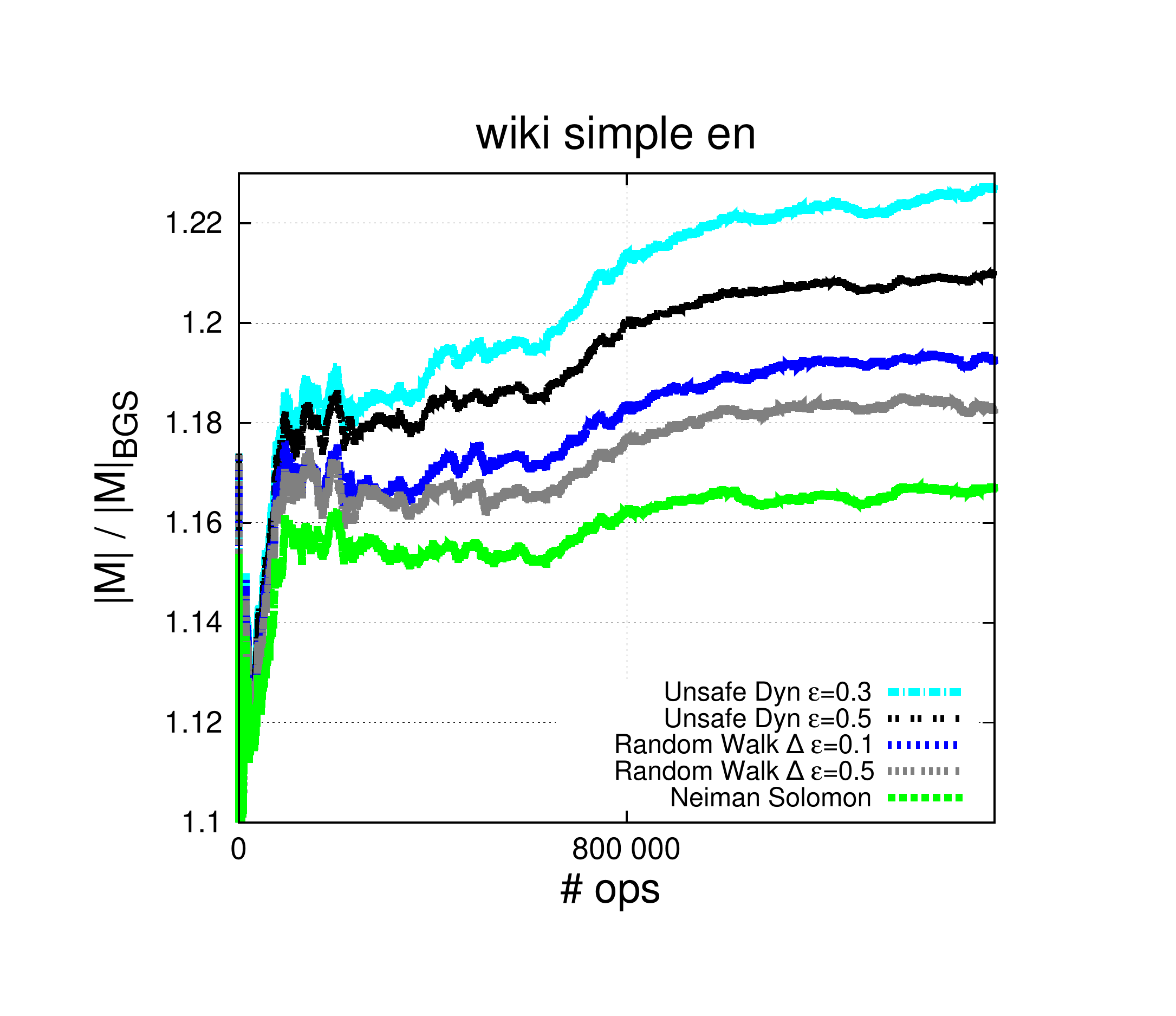}
\vspace*{-.75cm}
\caption{Matching size over time compared to Baswana, Gupta, Sen on the two real dynamic instances dewiki and wiki\_simple\_en.}
\vspace*{-.5cm}
\label{fig:real}
\end{figure}
\else
\fi{}
\ifTR In future work, it may be interesting to transfer results to the weighted case, and to combine our algorithms with simple data reductions rules such as \cite{DBLP:conf/esa/KorenweinNNZ18}. \ifTR It could be interesting to use these dynamic matching algorithms to derive dynamic multilevel algorithms  for example for graph partitioning~\cite{DBLP:conf/gecco/MoreiraP018,DBLP:conf/europar/Akhremtsev0018}. \fi{} Another direction will be to explore the parallelization potential of random walk-based algorithms. As real-world dynamic instances with insertions \emph{and} deletions are currently unavailable to the public, we plan to open a graph repository where we explicitly collect dynamic graphs.\fi{}
\renewcommand{\bibname}{\begin{flushleft} References \end{flushleft}}
\bibliographystyle{abbrv}
\bibliography{phdthesiscs}

\begin{thebibliography}{10}

\bibitem{2002:BGL:504206}
{\em The Boost Graph Library: User Guide and Reference Manual}.
\newblock Addison-Wesley Longman Publishing Co., Inc., Boston, MA, USA, 2002.

\bibitem{DBLP:conf/europar/Akhremtsev0018}
Y.~Akhremtsev, P.~Sanders, and C.~Schulz.
\newblock High-quality shared-memory graph partitioning.
\newblock In M.~Aldinucci, L.~Padovani, and M.~Torquati, editors, {\em Euro-Par
  2018: Parallel Processing - 24th International Conference on Parallel and
  Distributed Computing, Turin, Italy, August 27-31, 2018, Proceedings}, volume
  11014 of {\em Lecture Notes in Computer Science}, pages 659--671. Springer,
  2018.

\bibitem{ArarCCSW18}
M.~Arar, S.~Chechik, S.~Cohen, C.~Stein, and D.~Wajc.
\newblock Dynamic matching: Reducing integral algorithms to
  approximately-maximal fractional algorithms.
\newblock In {\em 45th International Colloquium on Automata, Languages, and
  Programming, {ICALP} 2018}, pages 7:1--7:16, 2018.

\bibitem{benchmarksfornetworksanalysis}
D.~Bader, A.~Kappes, H.~Meyerhenke, P.~Sanders, C.~Schulz, and D.~Wagner.
\newblock {Benchmarking for Graph Clustering and Partitioning}.
\newblock In {\em Encyclopedia of Social Network Analysis and Mining}.
  Springer, 2014.

\bibitem{BaswanaGS15}
S.~Baswana, M.~Gupta, and S.~Sen.
\newblock Fully dynamic maximal matching in ${O}(\log n)$ update time.
\newblock {\em {SIAM} J. Comput.}, 44(1):88--113, 2015.

\bibitem{berge57}
C.~Berge.
\newblock Two theorems in graph theory.
\newblock {\em Proceedings of the National Academy of Sciences},
  43(9):842--844, 1957.

\bibitem{Bernstein2016a}
A.~Bernstein and C.~Stein.
\newblock Faster fully dynamic matchings with small approximation ratios.
\newblock In {\em Proceedings of the 27th Symposium on Discrete Algorithms
  {SODA}}, pages 692--711. {SIAM}, 2016.

\bibitem{BhattacharyaCH17}
S.~Bhattacharya, D.~Chakrabarty, and M.~Henzinger.
\newblock Deterministic fully dynamic approximate vertex cover and fractional
  matching in {O(1)} amortized update time.
\newblock In {\em 19th International Conf. on Integer Programming and
  Combinatorial Optimization {IPCO}}, pages 86--98, 2017.

\bibitem{BhattacharyaHI18}
S.~Bhattacharya, M.~Henzinger, and G.~F. Italiano.
\newblock Deterministic fully dynamic data structures for vertex cover and
  matching.
\newblock {\em {SIAM} J. Comput.}, 47(3):859--887, 2018.

\bibitem{Bhattacharya2016}
S.~Bhattacharya, M.~Henzinger, and D.~Nanongkai.
\newblock New deterministic approximation algorithms for fully dynamic
  matching.
\newblock In {\em Proceedings of the 48th Annual Symposium on Theory of
  Computing}, pages 398--411. {ACM}, 2016.

\bibitem{Bhattacharya2017b}
S.~Bhattacharya, M.~Henzinger, and D.~Nanongkai.
\newblock Fully dynamic approximate maximum matching and minimum vertex cover
  in \emph{O}(log\({}^{\mbox{3}}\) \emph{n}) worst case update time.
\newblock In P.~N. Klein, editor, {\em Proceedings of the Twenty-Eighth Annual
  {ACM-SIAM} Symposium on Discrete Algorithms {SODA}}, pages 470--489. {SIAM},
  2017.

\bibitem{DBLP:conf/europar/BirnOSSS13}
M.~Birn, V.~Osipov, P.~Sanders, C.~Schulz, and N.~Sitchinava.
\newblock Efficient parallel and external matching.
\newblock In {\em Euro-Par 2013}, volume 8097 of {\em LNCS}, pages 659--670.
  Springer, 2013.

\bibitem{CharikarS18}
M.~Charikar and S.~Solomon.
\newblock Fully dynamic almost-maximal matching: Breaking the polynomial
  worst-case time barrier.
\newblock In {\em 45th International Colloquium on Automata, Languages, and
  Programming {ICALP}}, pages 33:1--33:14, 2018.

\bibitem{UFsparsematrixcollection}
T.~Davis.
\newblock {The University of Florida Sparse Matrix Collection,
  \url{http://www.cise.ufl.edu/research/sparse/matrices}, 2008}.

\bibitem{DBLP:journals/mp/DolanM02}
E.~D. Dolan and J.~J. Mor{\'{e}}.
\newblock Benchmarking optimization software with performance profiles.
\newblock {\em Math. Program.}, 91(2):201--213, 2002.

\bibitem{DH03a}
D.~Drake and S.~Hougardy.
\newblock {A Simple Approximation Algorithm for the Weighted Matching Problem}.
\newblock {\em Information Processing Letters}, 85:211--213, 2003.

\bibitem{DBLP:conf/alenex/DroschinskyMT20}
A.~Droschinsky, P.~Mutzel, and E.~Thordsen.
\newblock Shrinking trees not blossoms: {A} recursive maximum matching
  approach.
\newblock In {\em Proceedings of the Symposium on Algorithm Engineering and
  Experiments, {ALENEX} 2020}, pages 146--160. {SIAM}, 2020.

\bibitem{edmonds1965paths}
J.~Edmonds.
\newblock Paths, trees, and flowers.
\newblock {\em Canadian Journal of mathematics}, 17(3):449--467, 1965.

\bibitem{Gabow74}
H.~N. Gabow.
\newblock {\em Implementation of Algorithms for Maximum Matching on
  Nonbipartite Graphs.}
\newblock PhD thesis, Stanford University, Stanford, CA, USA, 1974.

\bibitem{GalilMG86}
Z.~Galil, S.~Micali, and H.~N. Gabow.
\newblock An {O(|E||V|} log {|V|)} algorithm for finding a maximal weighted
  matching in general graphs.
\newblock {\em {SIAM} Journal Computing}, 15(1):120--130, 1986.

\bibitem{DBLP:conf/soda/0001LSSS19}
F.~Grandoni, S.~Leonardi, P.~Sankowski, C.~Schwiegelshohn, and S.~Solomon.
\newblock {(1} + {\(\epsilon\)})-approximate incremental matching in constant
  deterministic amortized time.
\newblock In {\em Proceedings of the 20th Symposium on Discrete Algorithms},
  pages 1886--1898. {SIAM}, 2019.

\bibitem{GuptaP13}
M.~Gupta and R.~Peng.
\newblock Fully dynamic {(1+} e)-approximate matchings.
\newblock In {\em 54th Symposium on Foundations of Computer Science, {FOCS}},
  pages 548--557. {IEEE} Computer Society, 2013.

\bibitem{hopkarp71}
J.~E. Hopcroft and R.~M. Karp.
\newblock A {$n^{5/2}$} algorithm for maximum matchings in bipartite.
\newblock In {\em 12th Annual Symposium on Switching and Automata Theory
  ({SWAT})}, pages 122--125, 1971.

\bibitem{IvkovicL93}
Z.~Ivkovic and E.~L. Lloyd.
\newblock Fully dynamic maintenance of vertex cover.
\newblock In {\em 19th International Workshop Graph-Theoretic Concepts in
  Computer Science}, volume 790 of {\em LNCS}, pages 99--111, 1993.

\bibitem{DBLP:conf/esa/KorenweinNNZ18}
V.~Korenwein, A.~Nichterlein, R.~Niedermeier, and P.~Zschoche.
\newblock Data reduction for maximum matching on real-world graphs: Theory and
  experiments.
\newblock In {\em 26th European Symposium on Algorithms {ESA}}, volume 112 of
  {\em LIPIcs}, pages 53:1--53:13. Schloss Dagstuhl - Leibniz-Zentrum fuer
  Informatik, 2018.

\bibitem{DBLP:conf/www/Kunegis13}
J.~Kunegis.
\newblock {KONECT:} the koblenz network collection.
\newblock In L.~Carr, A.~H.~F. Laender, B.~F. L{\'{o}}scio, I.~King,
  M.~Fontoura, D.~Vrandecic, L.~Aroyo, J.~P.~M. de~Oliveira, F.~Lima, and
  E.~Wilde, editors, {\em 22nd World Wide Web Conference, {WWW} '13}, pages
  1343--1350. International World Wide Web Conferences Steering Committee /
  {ACM}, 2013.

\bibitem{snap}
J.~Lescovec.
\newblock Stanford {N}etwork {A}nalysis {P}ackage ({S}{N}{A}{P}).
\newblock \url{http://snap.stanford.edu/index.html}.

\bibitem{MauSan07}
J.~Maue and P.~Sanders.
\newblock {Engineering Algorithms for Approximate Weighted Matching}.
\newblock In {\em Proceedings of the 6th Workshop on Experimental Algorithms
  ({WEA'07})}, volume 4525 of {\em LNCS}, pages 242--255. Springer, 2007.

\bibitem{DBLP:books/cu/MehlhornN99}
K.~Mehlhorn and S.~N{\"{a}}her.
\newblock {\em {LEDA:} {A} Platform for Combinatorial and Geometric Computing}.
\newblock Cambridge University Press, 1999.

\bibitem{DBLP:conf/focs/MehtaSVV05}
A.~Mehta, A.~Saberi, U.~V. Vazirani, and V.~V. Vazirani.
\newblock Adwords and generalized on-line matching.
\newblock In {\em 46th {IEEE} Symposium on Foundations of Computer Science
  {(FOCS)}}, pages 264--273. {IEEE} Computer Society, 2005.

\bibitem{DBLP:conf/focs/MicaliV80}
S.~Micali and V.~V. Vazirani.
\newblock An ${O}(\sqrt{|{V}|} |{E}|)$ algorithm for finding maximum matching
  in general graphs.
\newblock In {\em 21st Symposium on Foundations of Computer Science}, pages
  17--27. {IEEE} Computer Society, 1980.

\bibitem{DBLP:conf/gecco/MoreiraP018}
O.~Moreira, M.~Popp, and C.~Schulz.
\newblock Evolutionary multi-level acyclic graph partitioning.
\newblock In H.~E. Aguirre and K.~Takadama, editors, {\em Proceedings of the
  Genetic and Evolutionary Computation Conference, {GECCO} 2018, Kyoto, Japan,
  July 15-19, 2018}, pages 332--339. {ACM}, 2018.

\bibitem{NeimanS16}
O.~Neiman and S.~Solomon.
\newblock Simple deterministic algorithms for fully dynamic maximal matching.
\newblock {\em {ACM} Trans. Algorithms}, 12(1):7:1--7:15, 2016.

\bibitem{OnakRubinfeld10}
K.~Onak and R.~Rubinfeld.
\newblock Maintaining a large matching and a small vertex cover.
\newblock In {\em STOC}, pages 457--464, 2010.

\bibitem{konect:unlink}
J.~Preusse, J.~Kunegis, M.~Thimm, T.~Gottron, and S.~Staab.
\newblock Structural dynamics of knowledge networks.
\newblock In {\em Proc. Int. Conf. on Weblogs and Social Media}, 2013.

\bibitem{Sankowski07}
P.~Sankowski.
\newblock Faster dynamic matchings and vertex connectivity.
\newblock In {\em SODA}, pages 118--126, 2007.

\bibitem{Solomon16}
S.~Solomon.
\newblock Fully dynamic maximal matching in constant update time.
\newblock In {\em 57th Symposium on Foundations of Computer Science {FOCS}},
  pages 325--334, 2016.

\bibitem{DBLP:books/daglib/0067705}
R.~E. Tarjan.
\newblock {\em Data structures and network algorithms}, volume~44 of {\em
  {CBMS-NSF} regional conference series in applied mathematics}.
\newblock {SIAM}, 1983.

\end{thebibliography}
\newpage
\begin{appendix}
\section{Instances}
\begin{table}[h!]
      \centering
      \caption{Basic properties of the benchmark set of static graphs obtained from~\cite{benchmarksfornetworksanalysis,UFsparsematrixcollection,snap}.}
      \begin{tabular}{lrr@{\hskip 13pt} lrr@{\hskip 13pt} }
      graph & $n$ & $m$ & graph & $n$ & $m$ \\
              \midrule

              \midrule
144 &\numprint{144649} & \numprint{1074393} &               eu-2005 &\numprint{862664} & \numprint{16138468} \\
3elt &\numprint{4720} & \numprint{13722} &                  fe\_4elt2 &\numprint{11143} & \numprint{32818} \\
4elt &\numprint{15606} & \numprint{45878} &                 fe\_body &\numprint{45087} & \numprint{163734} \\ 
598a &\numprint{110971} & \numprint{741934} &               fe\_ocean &\numprint{143437} & \numprint{409593} \\ 
add20 &\numprint{2395} & \numprint{7462} &                  fe\_pwt &\numprint{36519} & \numprint{144794} \\
add32 &\numprint{4960} & \numprint{9462} &                  fe\_rotor &\numprint{99617} & \numprint{662431} \\ 
amazon-2008 &\numprint{735323} & \numprint{3523472} &       fe\_sphere &\numprint{16386} & \numprint{49152} \\ 
as-22july06 &\numprint{22963} & \numprint{48436} &          fe\_tooth &\numprint{78136} & \numprint{452591} \\ 
as-skitter &\numprint{554930} & \numprint{5797663} &        finan512 &\numprint{74752} & \numprint{261120} \\
auto &\numprint{448695} & \numprint{3314611} &              in-2004 &\numprint{1382908} & \numprint{13591473} \\
bcsstk29 &\numprint{13992} & \numprint{302748} &            loc-brightkite\_edges &\numprint{56739} & \numprint{212945} \\
bcsstk30 &\numprint{28924} & \numprint{1007284} &           loc-gowalla\_edges &\numprint{196591} & \numprint{950327} \\
bcsstk31 &\numprint{35588} & \numprint{572914} &            m14b &\numprint{214765} & \numprint{1679018} \\
bcsstk32 &\numprint{44609} & \numprint{985046} &            memplus &\numprint{17758} & \numprint{54196} \\
bcsstk33 &\numprint{8738} & \numprint{291583} &             p2p-Gnutella04 &\numprint{6405} & \numprint{29215} \\
brack2 &\numprint{62631} & \numprint{366559} &              PGPgiantcompo &\numprint{10680} & \numprint{24316} \\
citationCiteseer &\numprint{268495} & \numprint{1156647} &  rgg\_n\_2\_15\_s0 &\numprint{32768} & \numprint{160240} \\
cnr-2000 &\numprint{325557} & \numprint{2738969} &          soc-Slashdot0902 &\numprint{28550} & \numprint{379445} \\
coAuthorsCiteseer &\numprint{227320} & \numprint{814134} &  t60k &\numprint{60005} & \numprint{89440} \\
coAuthorsDBLP &\numprint{299067} & \numprint{977676} &      uk &\numprint{4824} & \numprint{6837} \\
coPapersCiteseer &\numprint{434102} & \numprint{16036720} & vibrobox &\numprint{12328} & \numprint{165250} \\
coPapersDBLP &\numprint{540486} & \numprint{15245729} &     wave &\numprint{156317} & \numprint{1059331} \\
crack &\numprint{10240} & \numprint{30380} &                web-Google &\numprint{356648} & \numprint{2093324} \\
cs4 &\numprint{22499} & \numprint{43858} &                  whitaker3 &\numprint{9800} & \numprint{28989} \\ 
cti &\numprint{16840} & \numprint{48232} &                  wiki-Talk &\numprint{232314} & \numprint{1458806}\\
data &\numprint{2851} & \numprint{15093} &                  wing &\numprint{62032} & \numprint{121544} \\
email-EuAll &\numprint{16805} & \numprint{60260} &          wing\_nodal &\numprint{10937} & \numprint{75488} \\
enron &\numprint{69244} & \numprint{254449} &               wordassociation-2011 &\numprint{10617} & \numprint{63788} \\
      \bottomrule
      \end{tabular}
      \label{staticgraphs}
\label{tab:graphstable}
\end{table}
\vfill
\pagebreak
\begin{table}[h!]
      \centering
      \caption{Basic properties of the benchmark set of dynamic graphs with number of update operations $\mathcal{O}$. Most of the graphs only feature insertions. The only two expceptions are marked with a *. All of these graphs have been obtained from the KONECT graph database~\cite{konect:unlink}.} 
      \begin{tabular}{lrr@{\hskip 13pt}}
      graph & $n$ & $\mathcal{O}$ \\
\midrule
\midrule
amazon-ratings &\numprint{2146058} & \numprint{5838041} \\ 
citeulike\_ui &\numprint{731770} & \numprint{2411819} \\ 
dewiki$^*$ &\numprint{2166670} & \numprint{86337879} \\ 
dnc-temporalGraph &\numprint{2030} & \numprint{39264} \\ 
facebook-wosn-wall &\numprint{46953} & \numprint{876993} \\ 
flickr-growth &\numprint{2302926} & \numprint{33140017} \\ 
haggle &\numprint{275} & \numprint{28244} \\ 
lastfm\_band &\numprint{174078} & \numprint{19150868} \\ 
lkml-reply &\numprint{63400} & \numprint{1096440} \\ 
movielens10m &\numprint{69879} & \numprint{10000054} \\ 
munmun\_digg &\numprint{30399} & \numprint{87627} \\ 
proper\_loans &\numprint{89270} & \numprint{3394979} \\ 
sociopatterns-infections &\numprint{411} & \numprint{17298} \\ 
stackexchange-stackoverflow &\numprint{545197} & \numprint{1301942} \\ 
topology &\numprint{34762} & \numprint{171403} \\ 
wikipedia-growth &\numprint{1870710} & \numprint{39953145} \\ 
wiki\_simple\_en$^*$ &\numprint{100313} & \numprint{1627472} \\ 
youtube-u-growth &\numprint{3223590} & \numprint{9375374} \\
\bottomrule
\end{tabular}
\label{dyninstances}
\end{table}
\vfill
{}
{}

\end{appendix}
\end{document}